\documentclass[twocolumn,showpacs,preprintnumbers,prb,amsmath,amssymb,superscriptaddress ]{revtex4}

\usepackage{amsmath}    
\usepackage{amsfonts}  
\usepackage{amssymb}
\usepackage{graphicx}   
\usepackage{array}

\newcommand{\Tc}{$T_c$~}
\newcommand{\Hc}{$H_{c2}$~}
\newcommand{\Hczero}{$H_{c2}(0)$}

\newcommand{\etal}{{\it et al.}}
\newcommand{\eg}{{\it e.g.,}~}
\newcommand{\ie}{{\it i.e.,}~}

\newcommand{\BaPtx}{BaFe$_2$$_-$$_x$Pt$_x$As$_2$~}
\newcommand{\BaPt}{BaFe$_1$$_.$$_9$Pt$_0$$_.$$_1$As$_2$~}
\newcommand{\BaKx}{Ba$_1$$_-$$_x$K$_x$Fe$_2$As$_2$~}
\newcommand{\BaCox}{BaFe$_2$$_-$$_x$Co$_x$As$_2$~}
\newcommand{\BaNix}{BaFe$_2$$_-$$_x$Ni$_x$As$_2$~}
\newcommand{\KFeAs}{KFe$_2$As$_2$~}
\newcommand{\kzeroT}{$\kappa_{0}/T$~}

\begin{document}
  
\title{Isotropic multi-gap superconductivity in \BaPt from \\
thermal transport and spectroscopic measurements}

\author{Steven Ziemak}
\author{K.~Kirshenbaum}
\author{S.~R.~Saha}
\author{R.~Hu}
 \affiliation{Center for Nanophysics and Advanced Materials, Department of Physics, University of Maryland, College Park, MD 20742}
\author{J.-Ph.~Reid}
\author{R.~Gordon}
 \affiliation{D\'{e}partment de Physique \& RQMP, Universit\'{e} de Sherbrooke, Sherbrooke, Qu\'{e}bec, Canada J1K 2R1}
\author{L.~Taillefer}
 \affiliation{D\'{e}partment de Physique \& RQMP, Universit\'{e} de Sherbrooke, Sherbrooke, Qu\'{e}bec, Canada J1K 2R1}
 \affiliation{Canadian Institute for Advanced Research, Toronto, Canada M5G 1Z8}
\author{D.~Evtushinsky}
\author{S.~Thirupathaiah}
\author{S.~V.~Borisenko}
 \affiliation{Institute for Solid State Research, IFW-Dresden, P.O. Box 270116, D-01171 Dresden, Germany}
\author{A.~Ignatov}
\author{D.~Kolchmeyer}
\author{G.~Blumberg}
 \affiliation{Department of Physics and Astronomy, Rutgers, The State University of New Jersey, Piscataway, NJ 08854}
\author{J.~Paglione}
 \email{paglione@umd.edu}
 \affiliation{Center for Nanophysics and Advanced Materials, Department of Physics, University of Maryland, College Park, MD 20742}
 \affiliation{Canadian Institute for Advanced Research, Toronto, Canada M5G 1Z8}

\date{\today}

\begin{abstract}
Thermal conductivity, point contact spectroscopy, angle-resolved photoemission and Raman spectroscopy measurements were performed on BaFe$_1$$_.$$_9$Pt$_0$$_.$$_1$As$_2$ single crystals obtained from the same synthesis batch in order to investigate the superconducting energy gap structure using multiple techniques. 
Low temperature thermal conductivity was measured in the superconducting state as a function of temperature and magnetic field, revealing an absence of quasiparticle excitations in the $T \to 0$ limit up to 15~T applied magnetic fields. 
Point-contact Andreev reflection spectroscopy measurements were performed as a function of temperature using the needle-anvil technique, yielding features in the conductance spectra at both 2.5~meV and 7.0~meV scales consistent with a multi-gap scenario. 
Angle-resolved photoemission spectroscopy probed the electronic band structure above and below the superconducting transition temperature of $T_c$=23~K, revealing an isotropic gap of magnitude $\sim 3$~meV on both electron and hole pockets.
Finally, Raman spectroscopy was used to probe quasiparticle excitations in multiple channels, showing a threshold energy scale of 3~meV below $T_c$. 
Overall, we find strong evidence for an isotropic gap structure with no nodes or deep minima in this system, with a 3~meV magnitude gap consistently observed and a second, larger gap suggested by point contact spectroscopy measurements.
We discuss the implications that the combination of these results reveal about the superconducting order parameter in the \BaPtx doping system and how this relates to similar substituted iron pnictides.

\end{abstract}


\maketitle

\section{Introduction}

Since their discovery,\cite{Kamihara3296} the nature of the superconducting gap structure and pairing symmetry of iron-based superconductors (FeSCs) have attracted considerable attention.\cite{reviews} 
Of particular interest are the 122 iron pnictides, a class of intermetallic compounds with the ThCr$_2$Si$_2$ structure in which 
superconductivity is induced by several methods of chemical substitution. Aside from the more commonly studied substitution series of 3$d$ transition metal elements for iron in AFe$_{2-x}$T$_x$As$_2$ (A = Ba, Sr, Ca; T = Fe, Co, Ni...), previous work has shown that 5$d$ transition metal substitution also induces superconductivity, for example in the series \BaPtx where the maximum transition temperature \Tc approaches $\sim 25$~K.cite{Saha072204,Xhu104525}
Studies to date on this system have focused on crystal structure, resistivity, magnetic susceptibility, and specific heat, but more detailed studies of the pairing energy gap are lacking and would be useful for comparisons to the more heavily studied \BaCox system.

In understanding the electronic structure of superconductors it is essential to 
use a comprehensive experimental approach, as each measurement technique has its 
own unique limitations and opportunities for error. For example, angle-resolved 
photoemission spectroscopy (ARPES) is considered to be one of the most direct ways to 
measure the momentum-space structure of the superconducting energy gap $\Delta$, but in many cases ARPES studies on FeSCs have disagreed 
with bulk probes on the presence or absence of nodes in the order parameter. 
This apparent paradox may be a consequence of the surface-sensitive nature of the ARPES measurement or experimental limitations, where issues such as surface reconstruction, surface depairing, and/or resolution issues may impair the interpretation of experimental results.\cite{Hirschfeld124508} Other measurements designed to probe the gap structure similarly suffer from probe-dependent circumstances or are simply less direct probes of a gap's angular dependence. 
For this reason, we have taken the approach of comparing and contrasting four separate techniques: thermal conductivity, point-contact spectroscopy, ARPES, and Raman spectroscopy to investigate the \BaPtx system. Furthermore, we utilize samples for each experiment taken from the same synthesis batch, minimizing the error associated with comparisons of experimental results obtained from samples with varying quality, properties and origins, an issue particularly important in systems where chemical substitution is required.

Thermal conductivity measurements provide a powerful probe of electronic excitations in the zero temperature limit, when phonons have been frozen out. In a conducting solid at low temperatures, the temperature dependence of thermal conductivity $\kappa$ can be modeled as ${\kappa(T) = aT + bT^\alpha}$, where the $T$-linear term $aT$ represents the electronic contribution to heat conduction, while the ${T^\alpha}$ term represents the phonon contribution, where ${2\leq\alpha\leq3}$. By measuring ${\kappa/T}$ in the zero temperature limit the electron contribution can be isolated, given by $a\equiv\kappa_0/T$. It has been shown that the presence or 
absence of this residual thermal conductivity at zero magnetic field and its 
evolution in field can reveal the presence or absence of nodes or zeroes in the order parameter and provide additional information about gap structure. \cite{Taillefer483} For example, in the $d$-wave cuprate superconductors \kzeroT is nonzero at zero magnetic 
field.\cite{Proust147003} The presence of this same residual thermal conductivity has 
also been used as evidence for nodal $s$-wave or $d$-wave pairing in FeSCs such as 
\BaCox (Ref.~\onlinecite{Reid064501}) and \KFeAs (Ref.~\onlinecite{Reid087001}). Likewise, its absence has been taken as evidence for a nodeless $s$-wave gap structure in the 18~K superconductor LiFeAs (Ref.~\onlinecite{Tanatar054507}). As a function of magnetic field, the evolution of \kzeroT can provide further information about energy scale anisotropies or the presence of multiple gaps with different energy scales, and has been used to differentiate between $s$-wave superconductors with a single gap (or gaps of equal magnitude on multiple Fermi surface (FS) pockets) and those with multiple gaps of significantly different sizes, \eg differentiating doubly-gapped NbSe$_2$ from singly-gapped V$_3$Si (Ref.~\onlinecite{Boaknin117003}).

Another tool for probing electronic excitations is point-contact spectroscopy 
(PCS), also known as quasiparticle scattering spectroscopy. Measurements of 
electrical conductivity vs. DC bias voltage across a normal metal/superconductor 
junction provide an indirect probe of density of states, and fitting such 
spectra to the Blonder-Tinkham-Klapwijk (BTK) theory \cite{BTK} allows precise 
determination of gap size(s) as well as pairing symmetry.\cite{Daghero124509} 
PCS measurements on various doped FeSC systems have resulted in widely 
varying conclusions about their pairing symmetries, at times contradicting the 
results from other measurement techniques. Point-contact measurements in the 
Andreev reflection regime have shown evidence for a two-gap structure in \BaKx 
(Refs.~\onlinecite{Samuely507, Szabo012503}), while \BaCox spectra have been fit  
to both one- and two-gap $s$-wave models depending on the study and even the fitted features in a single spectrum \cite{Daghero124509,Samuely507} (\ie inclusion or exclusion of zero-bias or higher bias features). Finally, PCS measurements of \BaNix have been used to argue for a two-gap $s$-wave structure at low Ni concentrations with nodes or deep minima emergent above optimal doping. \cite{Ren2891}

\begin{figure}[!]
  \centering
  \resizebox{3.25in}{!}{\includegraphics{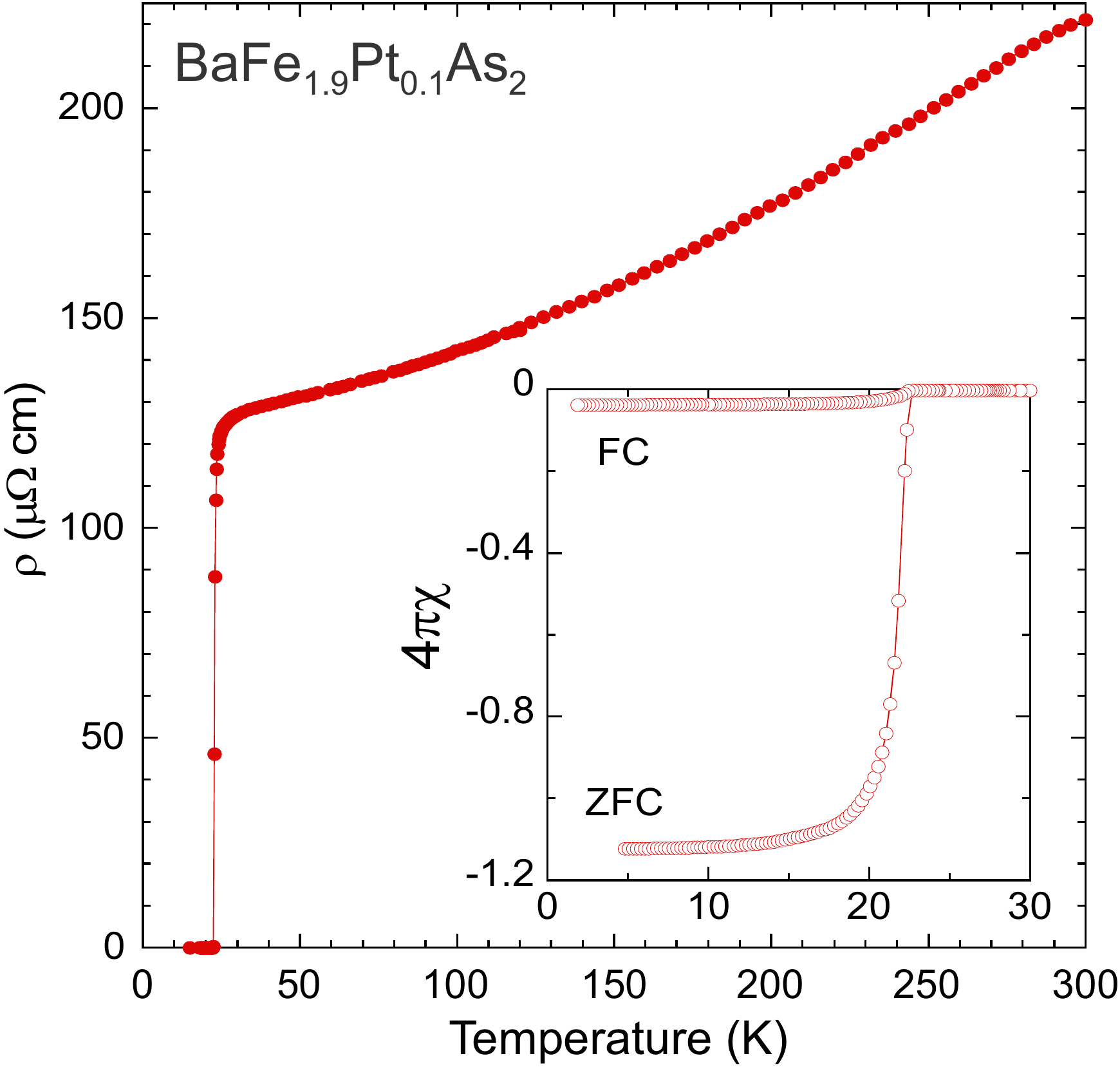}}
  \caption{\label{RT} Characteristic properties of the superconducting transition in single-crystal samples \BaPt with resistivity (main panel) and magnetic susceptibility (inset) measurements exhibiting sharp transition features consistent with a transition temperature at $T_c$=23~K, as observed in other bulk measurements.\cite{Saha072204} Magnetic susceptibility measurements performed in zero-field-cooled (ZFC) and field-cooled (FC) conditions with a magnetic field of 1~mT indicate a full 100\% 
  superconducting volume fraction (inset).}
\end{figure}

Angle-resolved photoemission spectroscopy (ARPES) provides a more direct picture 
of pairing symmetry and gap magnitude. In the case of 122 FeSCs, 
previous work has found two nodeless and nearly isotropic gaps of different 
sizes on various FS pockets in \BaKx (Ref.~\onlinecite{Ding47001}). ARPES measurements on \BaCox 
have shown evidence for hole pockets disappearing upon increasing Co substitution,\cite{Liu419} while a separate study focusing on gap structure \cite{Terashima7330} observed the presence of two isotropic gaps and suggested a connection between FS nesting and Cooper pairing based on comparison with the hole-doped \BaKx case. Finally, to date less extensive studies of Raman spectroscopy have been completed in the 122 FeSC materials,\cite{Daghero124509,Arham13047426}, where one study has shown band anisotropy in \BaCox supporting a nodal $s$-wave model.\cite{Muschler180510}

In this article, we report the results of four independent studies of the superconducting gap structure in the FeSC material \BaPt with $T_c$=23~K. We find consistent observations of an isotropic superconducting gap structure in this material using four techniques. 
Thermal conductivity measurements show no residual transport in the zero temperature limit under applied magnetic fields of up to 15~T, implying an isotropic, fully gapped $s$-wave structure with no nodes or deep minima.
PCS measurements reveal features in the conductance spectra consistent with a multi-gap scenario. Fitting spectra to a two-gap BTK theory shows that these features are consistent with the presence of two gaps of magnitude $\Delta_1$=2.5~meV and $\Delta_2$=7.0~meV.
ARPES measurements show a 3~meV energy gap in the density of states that develops below $T_c$ and whose size is independent of direction in $k$-space, confirming the isotropic nature of the smaller gap structure. 
Raman spectroscopy reveals excitations in the $A_{1g}$ and $B{_2g}$ channels 
with thresholds corresponding to a 3~meV gap size consistent with the other measurement techniques.

Details of sample growth and characterization are outlined in Sec.~II, followed by experimental results for thermal conductivity (Sec.~III), PCS (Sec.~IV), ARPES (Sec.~V) and Raman spectroscopy (Sec.~VI), and general discussion and comparisons in Sec.~VII followed by conclusions in Sec.~VIII.

\section{Sample Characterization}

Single crystals of \BaPt were grown from prereacted FeAs and PtAs powders and elemental Ba using the FeAs self-flux method 
\cite{Saha037005}, which yielded large crystals with typical dimensions 0.1 $\times$ 1 $\times$ 2 mm$^3$ (Ref.~\onlinecite{Saha072204}). Single crystal x-ray diffraction and Rietveld refinement determined a precise Fe:Pt ratio and an exact chemical formula of BaFe$_1$$_.$$_9$$_0$$_6$$_($$_8$$_)$Pt$_0$$_.$$_0$$_9$$_4$$_($$_8$$_)$As$_2$. Previous x-ray measurements have shown that Pt substitution reduces the $c$-axis length and $c/a$ ratio while increasing the $a$-axis length and unit cell volume compared to pure BaFe$_2$As$_2$ (Ref.~\onlinecite{Saha072204}). Characterization by energy- and wavelength-dispersive x-ray spectroscopy verified the same substitutional concentrations across several specimens.

As shown in Fig.~1, resistivity measurements exhibit the expected metallic behavior 
and a sharp transition to the superconducting state with an onset of $T_c$=23~K and zero resistance achieved by 21.5~K for a transition width of $\Delta$\Tc $<$ 1.5 K. No kink is observable at higher temperatures that would indicate a magnetically ordered phase, as is seen in the parent compound and underdoped samples.\cite{Xhu104525,Kirsh144518} Magnetic field suppresses \Tc down to approximately 16~K in a 14~T field (not shown), consistent with an upper critical field of approximately 45~T as determined by fits to the Werthamer-Helfand-Hohenberg approximation as demonstrated previously.\cite{Saha072204} However, \Hc(0) may be as high as ~65 T based on a linear fit to \Hc($T$), which would be consistent with similar materials such as SrFe$_2$$_-$$_x$Ni$_x$As$_2$ (Ref.~\onlinecite{Butch024518}).

DC magnetic susceptibitity was measured with field applied along the $ab$-plane, under zero-field-cooled conditions and field~cooled with a 1~mT applied field. As shown in the inset of Fig.~1, a sharp transition is observed at $T_c$=23~K into 
the diamagnetic state with a full-volume shielding fraction observed, as indicated by {$4\pi\chi = -1$} by 17 K. Previous specific heat measurements\cite{Saha072204} have confirmed the bulk nature of the transition, with a jump in {$C_p(T)$} that occurs slightly below \Tc. The size of the jump was estimated as {$\Delta$}{$C_p$/\Tc} $\approx$ 20~mJ~mol$^{-1}$K$^{-1}$, and assuming the BCS weak-coupling ratio of {$\Delta$$C_p$/$\gamma$\Tc}=1.43 applies yields a value for {$\gamma = 16(2)$}~mJ~mol$^{-1}$K$^{-2}$.\cite{Saha072204}.

\section{Thermal Conductivity}

\begin{figure}[!]
  \centering
 \includegraphics[width = 3.25 in]{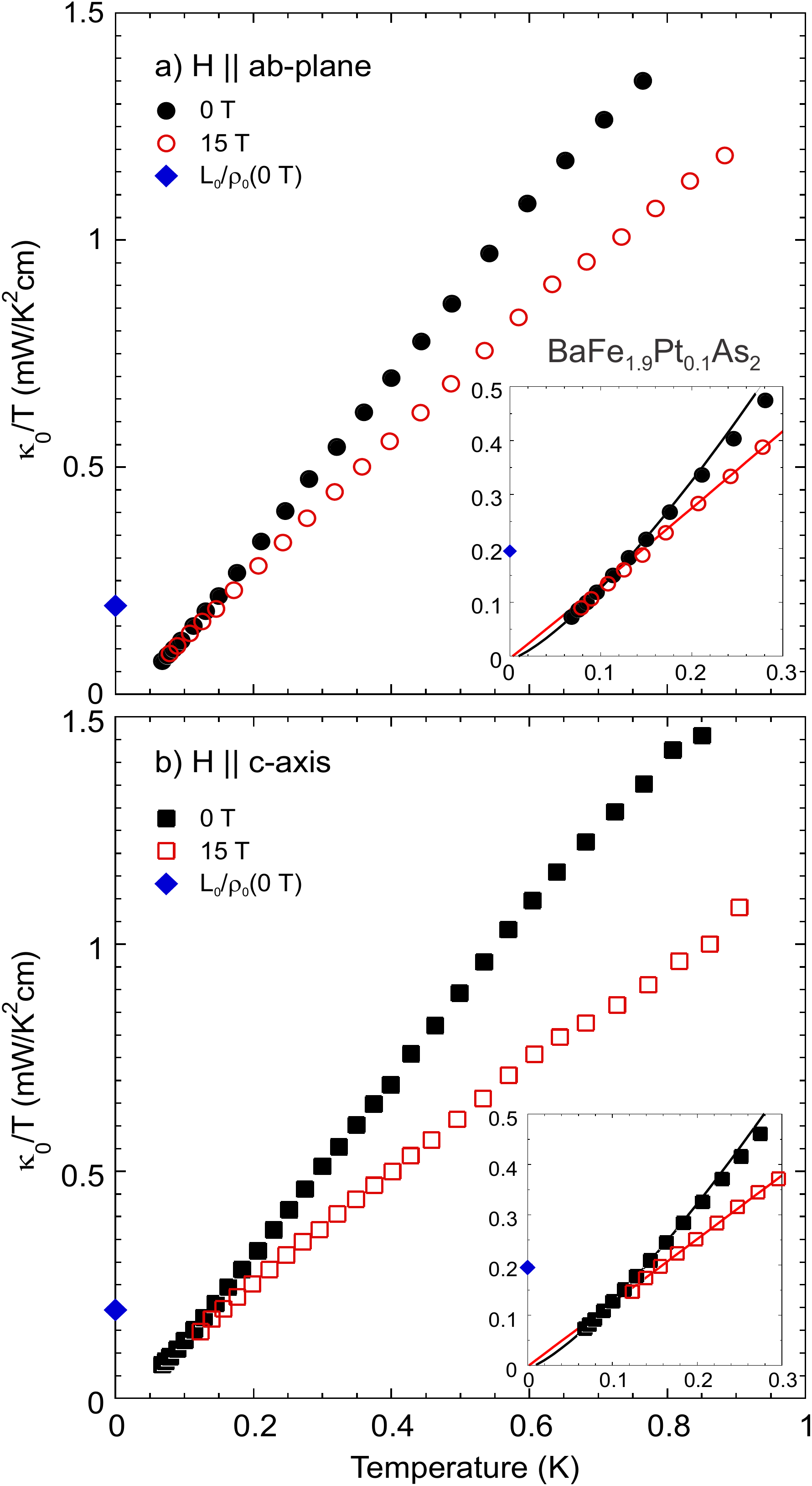}
 \caption{\label{Kappa-T} Low temperature thermal conductivity of a crystal of \BaPt measured in zero (black, closed symbols) and 15~T (red, open symbols) applied magnetic fields, oriented along a) the crystallographic $ab$ basal plane (circles) and b) the $c$-axis direction (squares). Zero field measurements were repeated in each orientation using the same crystal for direct comparison. Insets show a zoom of the low-temperature detail, with solid lines representing power law fits of the form $\kappa/T = a+bT^{\alpha-1}$ to the data points below 150 mK. Blue diamonds in all panels represent the normal state electronic thermal conductivity estimated using the Wiedemann-Franz law (see text).}
\end{figure}

Thermal conductivity was measured using a one-heater, two-thermometer steady state technique as a function of temperatures down to 60~mK in a dilution refrigerator. 
Temperature sweeps were repeated in different fixed magnetic field values from 0 T to 
15 T, applied in both parallel ($H\parallel ab$) and perpendicular ($H\parallel c$) orientations with respect to the crystallographic basal plane. Measured data, presented as $\kappa/T$ vs. $T$ and shown in Fig.~2, were fit to the form $\kappa/T = a+bT^{\alpha-1}$ in order to extrapolate to the $T\to 0$ limit and obtain the residual electronic term $\kappa_0/T$. 

\begin{figure}[!]
  \centering
 \includegraphics[width = 3.25 in]{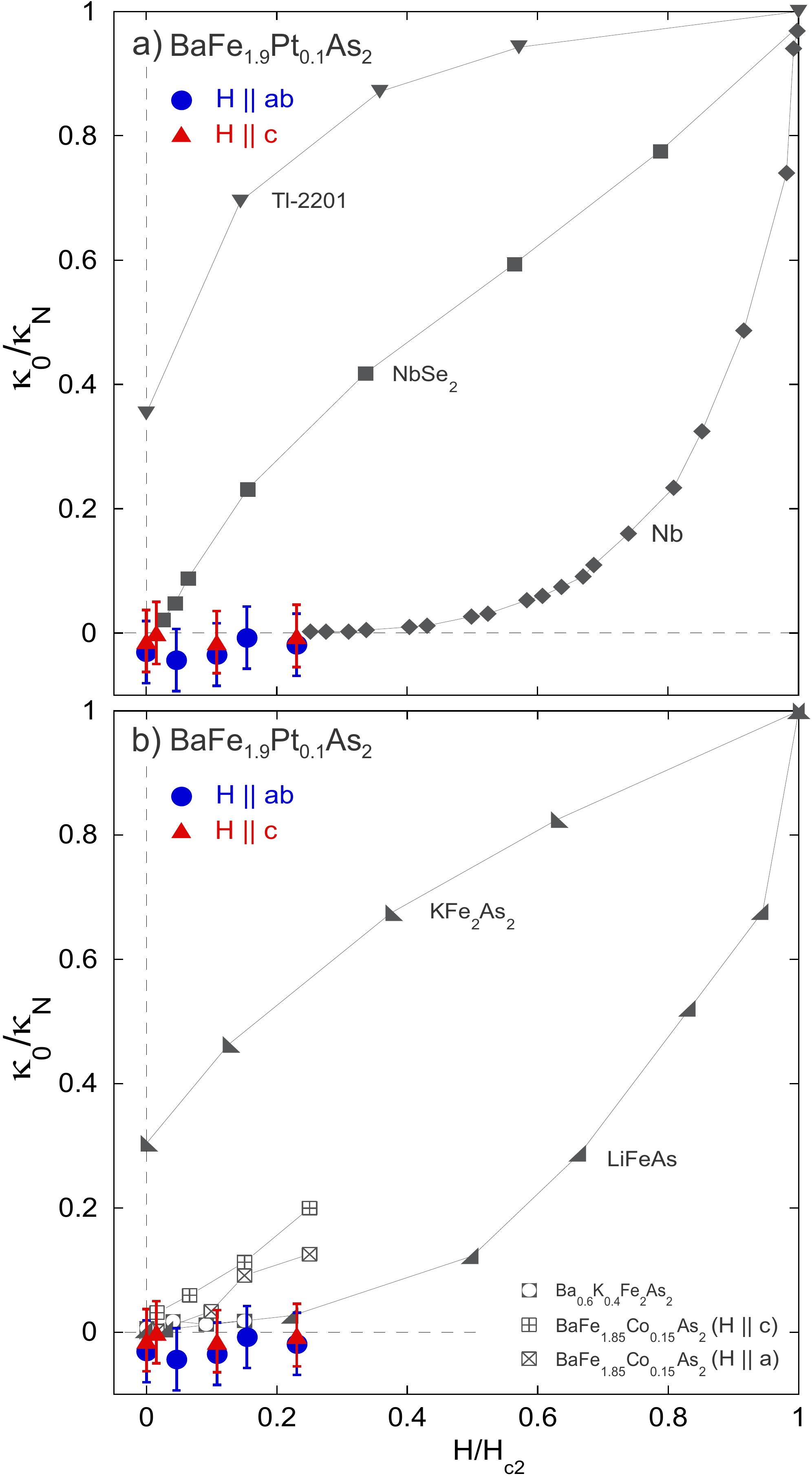}
 \caption{\label{Kappa-H} Residual thermal conductivity $\kappa_0/T$ of \BaPt extracted from temperature sweep measurements and normalized to the normal state conductivity $\kappa_N/T$, plotted as a function of reduced magnetic field and compared to several characteristic superconductors. 
 As shown in panel a), the lack of increase in $\kappa_0/\kappa_N$ in \BaPt is comparable to the behavior of isotropic, single-gapped $s$-wave superconductors such as elemental Nb (Ref.~\onlinecite{Lowell65}) that exhibit an exponentially slow activated increase with field, and is in contrast to the cases for multi-band $s$-wave superconductor NbSe$_2$ (Ref.~\onlinecite{Boaknin117003}) and $d$-wave nodal superconductor Tl$_2$Ba$_2$CuO$_{6+\delta}$ (Ref.~\onlinecite{Proust147003}), which exhibit a much faster rise with field due to nodal or small-gap low energy quasiparticle excitations. Panel b) compares \BaPt to the range of behaviors found in iron-based superconductors such as KFe$_2$As$_2$ (Ref.~\onlinecite{Reid084013}), Ba$_{0.6}$K$_{0.4}$Fe$_2$As$_2$ (Ref.~\onlinecite{Reid084013}), BaFe$_{1.85}$Co$_{0.15}$As$_2$ (Ref.~\onlinecite{Reid064501}), and LiFeAs (Ref.~\onlinecite{Tanatar054507}) as noted.}
\end{figure}

The residual electronic contribution $\kappa/T$ clearly approaches a value of zero as $T\to0$, indicating a lack of zero 
energy quasiparticles in the superconducting state. (The 
fitted value of $\kappa_{0}/T$ is actually slightly negative, but 
is within experimental error of zero). As a point of comparison, one can calculate the expected $T\to 0$ limit electronic contribution to thermal conductivity from the Wiedemann-Franz law, ${\kappa_0/T = L_0/\rho_0}$, where $L_0 = (\pi^2/3)(k_B/e)^2$=$2.44$$\times$$10^{-8}$~W~$\Omega$~K$^{-2}$. Using the estimated residual zero-temperature resistivity value of $\rho_0$=125~$\mu\Omega$~cm based on extrapolating the zero-field data (c.f. Fig.~1) yields an approximate value of $L_0/\rho_0$=0.195~mW~K$^{-2}$~cm$^{-1}$, shown in Fig.~2 as blue diamonds on the $T$=0 axis.   As evident in Fig.~2, this estimated normal state limit is many orders of magnitude larger than the maximum possible extrapolated value of $\kappa_{0}/T$ allowed by error, suggesting a fully gapped superconducting order parameter with no nodes or deep minima. This is to be contrasted with the finite $\kappa_{0}/T$ expected for superconductors with either symmetry-imposed or accidental nodes in their gap structure, such as observed in the $d$-wave superconductor Tl$_2$Ba$_2$CuO$_{6+\delta}$ (Tl-2201) \cite{Proust147003} or in $c$-axis transport measurements of \BaCox away from optimal doping\cite{Reid064501}, respectively. 

The absence of any low-energy excitations is also evident in magnetic field measurements up to 15~T, or approximately 25\% of \Hczero, as shown in Fig.~2 for both $H\parallel ab$ and $H\parallel c$ orientations in panels a) and b), respectively, and summarized as a function of field in Fig.~3.
As a function of increasing magnetic field, it has been shown that single-gap isotropic $s$-wave superconductors such as Nb (Ref.~\onlinecite{Lowell65}) continue to lack low-energy quasiparticle excitations with increasing magnetic field until vortex core bound states begin to delocalize, a process that proceeds exponentially with the ratio of intervortex spacing to coherence length (or essentially the magnitude of magnetic field) up to the uppper critical field.
Multiband $s$-wave superconductors such as NbSe$_2$ also lack any \kzeroT term in zero field due to lack of nodes in the gap structure. However, the vortex core delocalization process can be accelerated in such cases due to the presence of reduced gap magnitudes on one or more bands, resulting in an onset of low-energy excitations and a finite and increasing value of $\kappa_0/T$ at fields much smaller than $H_{c2}$.\cite{Boaknin117003}. 

The evolution of $\kappa_0/T$ with magnetic field in \BaPt is compared to the behavior of several characteristic superconductors in Fig.~3a), and to varying behaviors observed in the FeSC family in Fig.~3b). The lack of any increase in residual term in \BaPt is made even more clear when plotted as a function of field normalized to \Hczero ~(using 65~T for \BaPt as a conservative estimate): the evolution in \BaPt is comparable to that of elemental Nb (Ref.~\onlinecite{Lowell65}), and is in stark contrast to that of the $d$-wave and multiband $s$-wave superconductors Tl-2201 and NbSe$_2$, respectively. Within the FeSC family, such a flat response is not unprecedented, and is in fact favorably comparable to the full isotropic gap scenarios deduced for both Ba$_{0.6}$K$_{0.4}$Fe$_2$As$_2$ (Ref.~\onlinecite{Reid084013}) and LiFeAs (Ref.~\onlinecite{Tanatar054507}).

In the case of a multi-band scenario, which is the case for \BaPt as measured by ARPES (see Sect.~V), the suppression of the maximum energy gap on portions of the band structure can lead to an enhancement of tunneling between vortex core bound states, resulting in an effectively reduced critical field for those portions. In the case of NbSe$_2$, this field scale is identified with a shoulder in $\kappa_0/\kappa_N(H)$ near $H^*\sim H_{c2}/9$, consistent with an energy gap ratio $\Delta_{min}/\Delta_{max} \simeq 1/3$ as follows from the fact that $H_{c2} \propto \Delta^2/v_F$$^2$, where $v_F$ is the Fermi velocity.\cite{Boaknin117003}
In comparison, in \BaPt the lack of any rise whatsoever in $\kappa_0/\kappa_N(H)$ in the field range studied  up to $H_{c2}/4$ suggests a much weaker reduction from the maximum gap magnitude on those portions of the Fermi surface with a smaller gap, assuming the same assumptions apply.
However, as discussed for the case of LiFeAs (Ref.~\onlinecite{Tanatar054507}), the lack of any signature of a rise in $\kappa_0/\kappa_N(H)$ cannot completely rule out the existence of a smaller gap in the band structure, whose presence may be compatible these observations if for instance the conductivity contribution of the small-gap Fermi surface is much smaller that that of the large-gap component, or if coherence lengths are comparable due to scaling of Fermi velocities (\ie since $\xi_0 \sim v_F/\Delta$). In fact, the case of LiFeAs, as will be discussed below, is a pertinent comparison for the multi-band (and multi-gap) nature of \BaPt because recent ARPES experiments on LiFeAs do indeed find indications of two gap magnitudes of approximately 2-3~meV and 5-6~meV (Refs.~\onlinecite{Borisenko67002,Allan563,Umezawa37002}), with mild anisotropies reported but with gap minima not reaching below $\sim$2~meV. Therefore, it is likely that \BaPt also shares the situation of having similar coherence lengths across bands, making determination of the gap anisotropies with thermal conductivity difficult and complementary experiments necessary for further understanding.

\section{Point-Contact Spectroscopy}

Point-contact Andreev reflection spectroscopy (PCS) was used to further 
investigate the gap structure. Needle-anvil junctions were prepared using a 
sharpened Au tip attached to a specialized probe with a moveable stage, which 
was used to create contact between the tip and the sample. The samples used were 
large platelets extracted from FeAs flux and were freshly cleaved before 
applying contacts and inserting them into the probe. The junctions whose results are presented below had typical DC junction resistances on the order of 10-20 $\Omega$. AC resistance was measured against DC bias voltage with a lock-in amplifier used to stabilize the AC signal, over a range of temperatures using a $^4$He dipper cryostat.

\begin{figure}[!]
  \centering
 \includegraphics[width = 3.25in]{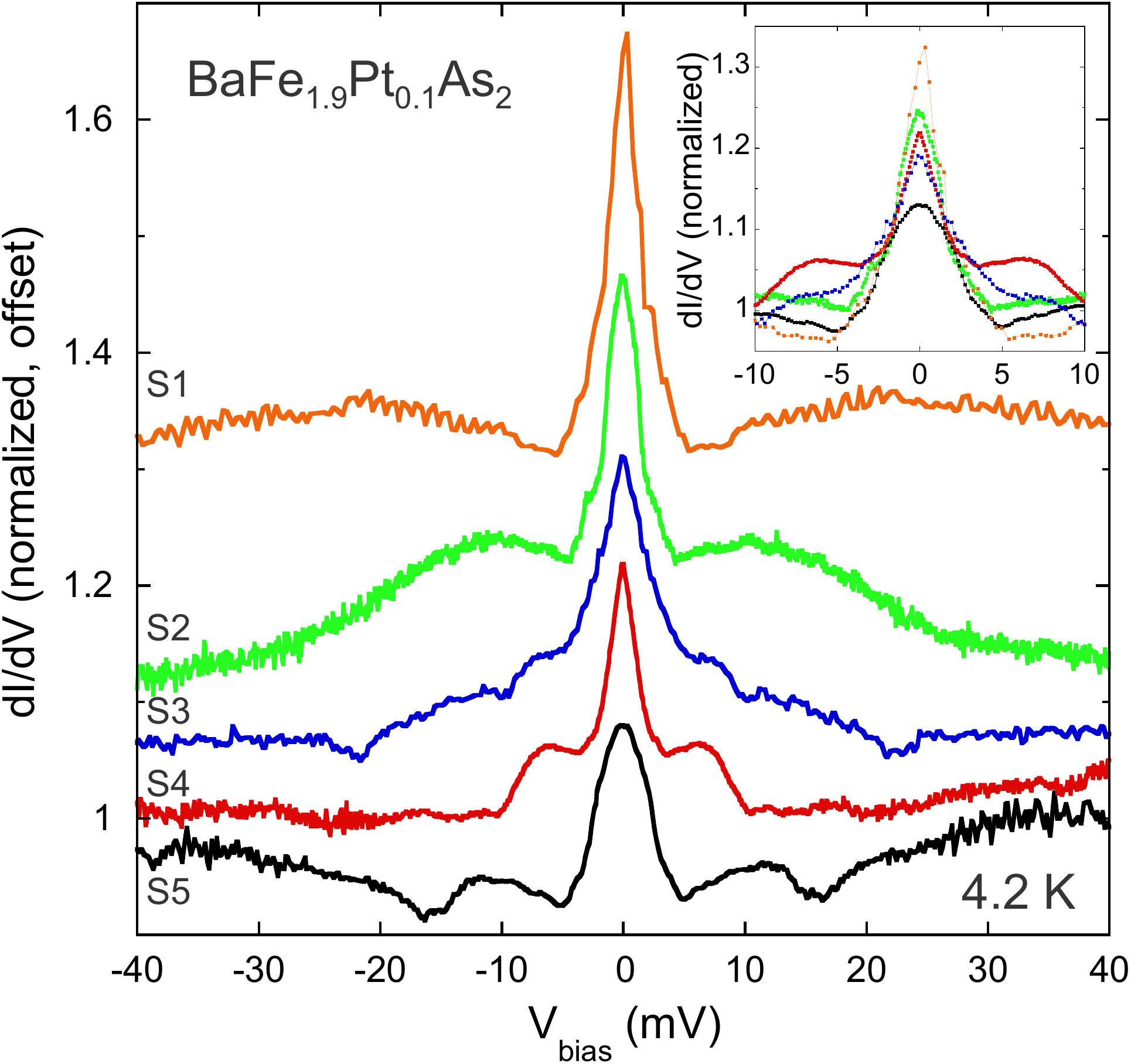}
 \caption{\label{mult_samples} Conductance curves obtained from point contact spectroscopy measurements performed on five \BaPt samples from the same synthesis batch, offset vertically for clarity. Each has a strong zero bias enhancement of similar width and either a depression or secondary enhancement at higher bias. These curves represent raw data without background  subtraction, and each curve is normalized to their high bias values. The inset presents the conductance curves for all five samples plotted without an offset in order to show the overlap of the gross gap features. The green curve (S2) corresponds to the data analyzed in Figs. 5 and 6.}
\end{figure}

In a normal metal-superconductor junction, measurements of the current-voltage ($I$-$V$) relation can reveal information about the nature of the superconducting gap(s) if the size of the junction is small compared to the superconductor's mean free path. Electrons incident on the junction from the normal metal with an energy of less than $\Delta$ can either be reflected back or generate a Cooper pair in the superconductor and reflect a hole back into the normal metal to conserve charge, spin, and momentum in a process known as Andreev reflection.\cite{Andreev1228} In a measurement of conductivity $dI/dV$ vs. bias voltage across the junction, the Andreev reflection process yields an enhancement of conductance when the chemical potential difference is less than $\Delta$. For the case of complete Andreev reflection in an isotropic $s$-wave superconductor, this enhancement occurs as a nominal doubling of the conductance that abruptly onsets at the gap energy. In the case of a junction with a large potential barrier between the normal metal and superconductor, conductivity is instead suppressed due to enhanced reflection at the junction, yielding the typical tunneling conductance spectrum.

For an isotropic $s$-wave superconductor, Andreev reflection leads to a flat doubling of the conductance from zero bias up to $\pm \Delta$,\cite{Andreev1228, BTK} while in other cases the conductance can exhibit sharper and larger enhancements near or at zero bias, in particular in nodal superconductors which exhibit sharper, non-flat peaks in their $dI/dV$ spectra. However, it is known that zero-bias conductance peaks can be observed in point contact measurements due to a wide variety of phenomena. Thus, consistency and 
replicability are crucial in verifying that observed $dI/dV$ features are indeed intrinsic and not artifacts of a measurement problem. 

We have measured conductance curves using junctions created with several different \BaPt crystals obtained from the same synthesis batch. A sampling of these $dI/dV$ curves are shown in Fig.~4, presented as raw data without any background normalization (see below) but only normalized to their individual high-bias values (\ie at 40~meV for each curve) in order to account for differences in junction resistance. 
Disregarding the slight differences in the shape of the 
curves, which results from differing background or scattering ($\gamma$) contributions, each of the $dI/dV$ curves in Fig.~4 features an abrupt low-bias peak with 
a width of approximately 2-3 meV, consistent with Andreev reflection in the superconducting state. Furthermore, all curves exhibit depressions or enhancements in conductance at higher energies (\ie closer to 5-6~meV). This overlap is emphasized by the inset of Fig.~4, which presents the same data without vertical offsets. 

Fig.~5 presents the temperature evolution of PCS conductance spectra for sample S2, now normalizing the $dI/dV$ data to the normal state spectra to remove the background contribution to conductance (c.f. Fig.~4). This is done by dividing out a polynomial fit to the data measured at 18~K from each temperature data set. Upon cooling below $T_c$ (Ref.~\onlinecite{PCSfootnote}), the features described above clearly emerge and evolve with decreasing temperature to reveal a sharp conductance enhancement and a depression at higher bias, suggesting features with at least two energy scales evolving in an order parameter-like fashion.

To better understand the features observed in the conductance spectra, we employ 
fitting to the Blonder-Tinkham-Klapwijk (BTK) model, which includes several parameters which can be used to fit the experimental data.\cite{BTK} The $I-V$ characteristics of 
a normal metal-superconductor (N-S) junction are determined by the gap size $\Delta$, a 
unitless barrier strength $Z$, and an inelastic scattering energy $\gamma$. The value of $Z$ determines the degree of tunneling across the junction: $Z$ = 0 indicates complete Andreev reflection, while $Z\to \infty$ indicates the tunneling limit. In the low-$Z$ case, the $dI/dV$ vs. $V_{bias}$ curve contains a peak at zero bias extending out to bias voltages corresponding to approximately +$\Delta$ and 
-$\Delta$, whereas for large $Z$, $dI/dV$ is suppressed at zero bias with enhancements
at $\pm$$\Delta$ due to a tunneling gap. The $\gamma$ term generally changes the shape of the $dI/dV$ curves and is therefore also critical in obtaining a good fit to measured data. In the case of a two-gap BTK model, the fit parameters include two gap sizes, $\Delta_1$ and $\Delta_2$, two barrier strengths, $Z_1$ and $Z_2$, two scattering energies, $\gamma_1$ and $\gamma_2$, and an additional weight factor $w$, which indicates the proportion of electrons incident on the barrier that interact with either gap. The composite two-gap BTK curve is simply a weighted average of two single-gap curves.

Fitting the conductance spectra features described above yields a good fit to the isotropic two-gap BTK formalism, as shown in Fig.~6 for sample S2 at 4.2~K. In particular, the fit captures both low- and high-bias features of enhancement and depression of conductance, respectively, via two sets of parameters: the low-bias conductance enhancement of this sample can be modeled by the existence of a nodeless gap with energy scale $\Delta_1$ = 2.5~meV, while the high bias depression is well modelled by including 
a second gap of magnitude $\Delta$$_2$ = 7.0~meV such that the fit closely matches the observed spectrum. These features are strikingly similar to those observed in PCS measurements performed on the two-gap superconductor MgB$_2$, which also exhibit peak-dip features that were understood in terms of a two-gap scenario with similar gap energies (but of course a much higher \Tc of 40~K).\cite{Naidyuk238}

\begin{figure}[!]
  \centering
 \includegraphics[width = 3.25 in]{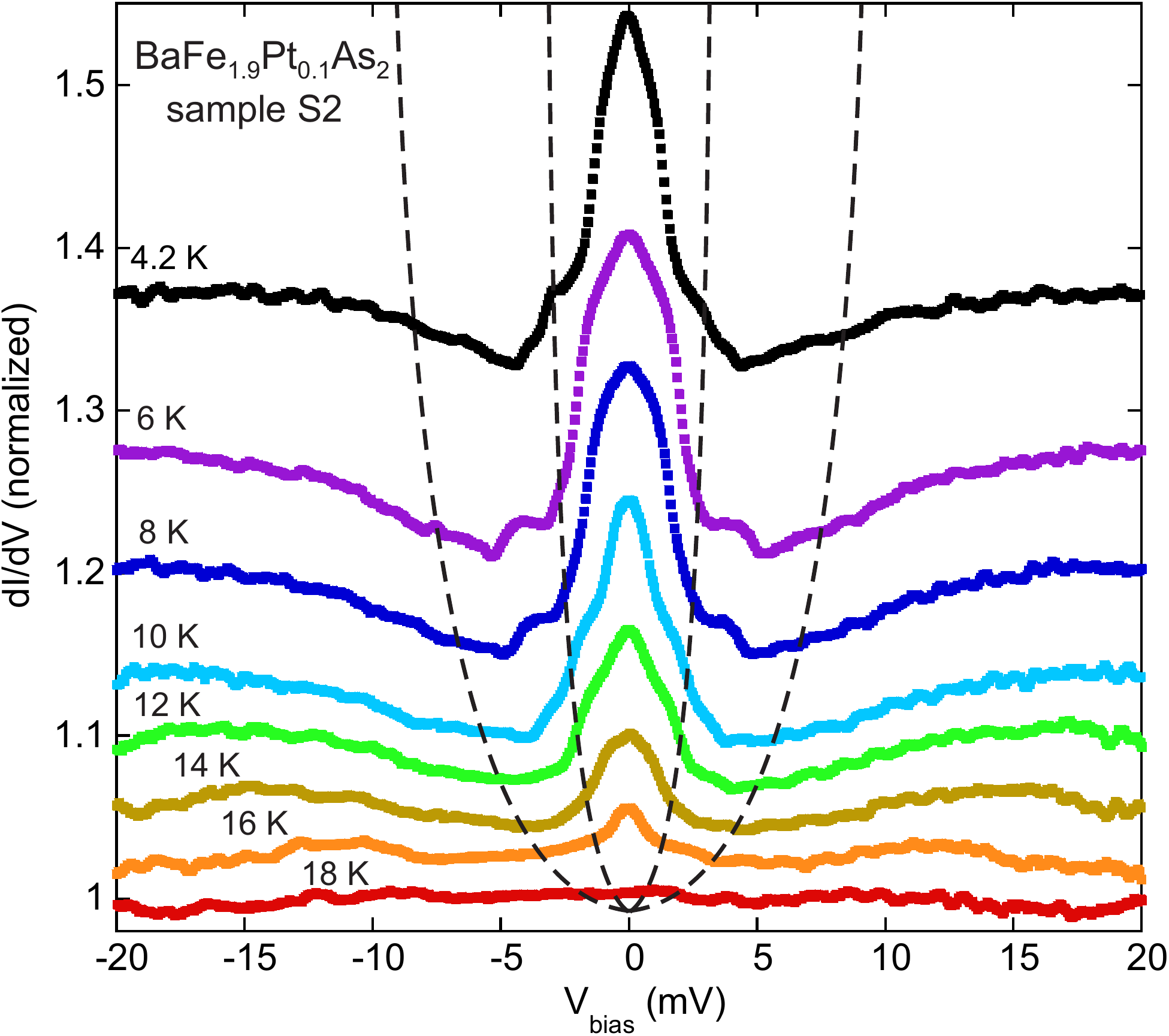}
 \caption{\label{dIdV} Normalized $dI/dV$ spectra for single crystal \BaPt with 
 high temperature background conductance divided out over a range of temperatures. Curves are offset for  clarity. Dashed lines serve as a guide to the eye to demonstrate that the two superconducting gap features, the low-bias conductance enhancement and higher bias 
 wells, decrease as expected up to \Tc (see text for details).}
\end{figure}

Interestingly, significant differences in the fit parameters in terms of the $Z$ parameter and broadening term $\gamma$ were necessary to fit the PCS spectra for \BaPt. For instance, the parameter set for the smaller gap spectrum ($\Delta$$_1$= 2.5 meV, $Z_1$ = 0, $\gamma$$_1$ = 1.0~meV) indicates complete Andreev reflection and a small broadening term, which is typical for such a transparent gap. However the larger gap spectrum ($\Delta$$_2$ = 7.0 meV , $Z_2$ = 7.0, $\gamma$$_2$ = 7.0 meV) appears closer to the tunneling limit with a larger $Z$ and exhibits much greater broadening. The weight factor of $w$=0.6 indicates comparable amounts of Andreev reflection and tunneling into each gap with more conduction into the smaller gap. 

While it is not clear why such significant differences were observed in 
the modeled barrier heights and scattering strengths of the two gaps, such differences are not unprecedented. For example, a two-gap model whose larger gap also has a larger $Z$ 
was observed in \BaKx (Ref.~\onlinecite{Samuely507}). 
The unusually large scattering strength deduced by the BTK fit for the larger gap, with $\gamma_2 \simeq \Delta_2$, is also unusual. However, it is in fact not unreasonable considering the high quasiparticle scattering rates deduced from pair-breaking experiments.\cite{Tanatar054507,Kirshenbaum140505} Kirshenbaum \etal~investigated pair-breaking 
scattering in a range of 122 iron pnictide materials with transition metal 
substitutions at the Fe site.\cite{Kirshenbaum140505} Using their observation of universal pair-breaking behavior in these compounds, one can estimate the transport scattering rate for \BaPt based on a \Tc value of 23~K, yielding $\Gamma \simeq$ 2.5 $\times$ 10$^{13}$ s$^{-1}$. 
This translates to an energy scale of $\hbar \Gamma \approx$ 16~meV, which is actually higher but of the same order as $\gamma_2$, suggesting there is a more profound question about strong scattering in the FeSC materials as addressed elsewhere.\cite{Kirshenbaum140505}. Alternatively, the contrast between $Z_1$ = 0 and $Z_2$ = 7.0 can find an explanation in terms of a relatively larger ``impedance mismatch'' between the tip and the large-gap band than for the small-gap band. Following the argument from thermal conductivity measurements on LiFeAs (Ref.~\onlinecite{Tanatar054507}) in regard to a mismatch in Fermi velocities, it is possible that the observation of this large contrast in $Z$ values and the (lack of) evidence for multi-band effects in thermal conductivity are not inconsistent.

\begin{figure}[!]
  \centering
 \includegraphics[width = 3.25in]{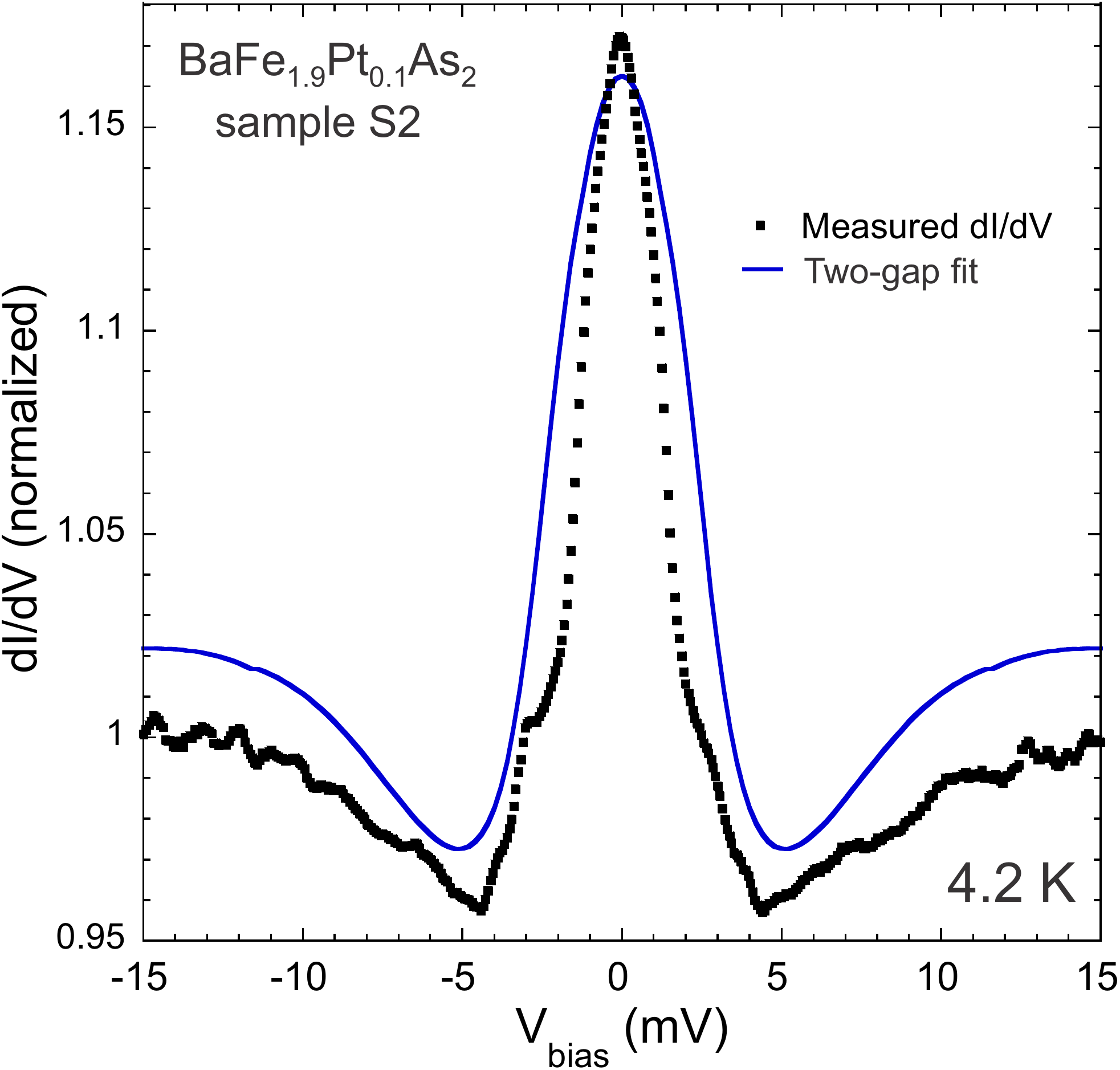}
 \caption{\label{4Kfit} Point contact conductance spectrum of \BaPt sample S2, measured at 4.2~K and normalized to remove the normal state background conductance (see text). Data are fit to an isotropic $s$-wave two-gap Blonder-Tinkham-Klapwijk model (see text for details), yielding two sets of parameters for a small-gap spectrum ($\Delta$$_1$= 2.5 meV, $Z_1$ = 0, $\gamma$$_1$ = 1.0~meV) and a large-gap spectrum ($\Delta$$_2$ = 7.0 meV , $Z_2$ = 7.0, $\gamma$$_2$ = 7.0 meV), with a weighting factor of $w$=0.6.}
\end{figure}

Finally, the appearance of spectral features at energy scales much larger than the predicted weak-coupling BCS gap value for a 23~K superconductor, $\Delta_{BCS}$ = 1.76${k_BT_c}$ = 3.5~meV, is puzzling but also not unprecedented in tunneling and PCS studies. 
For normal-superconductor junctions outside the ballistic limit, it has been 
shown that sharp dips can occur at higher bias voltage.\cite{Sheet134507} A conservative estimate gives an electron mean free path on the order of nanometers,\cite{Lu054009} which would place our micron-size junctions (assuming a perfect and full contact) in the diffusive rather than ballistic regime, and outside the Sharvin limit. 
However, as evidenced in Fig.~4 all sample spectra exhibit features at higher bias voltages, including both gap-like dips and Andreev-like shoulders. It is true that many PCS experiments observe dominant spectral features in a lower-energy range consistent with our prominent features near $\Delta_1$, such as in BaFe$_{1.86}$Co$_{0.14}$As$_2$ (Ref.~\onlinecite{Samuely507}) and in several 122's that exhibit a universal 2$\Delta$/$k_B$\Tc ratio of 3.1 (Ref.~\onlinecite{Zhang020515}). In the case of LiFeAs, observations of a small gap of 1.6~meV ($2\Delta/k_BT_c = 2.2$) have been reported from PCS measurements using Pb and Au tips \cite{Zhang94521}, which do not observe features associated with any larger gap as seen in ARPES experiments.\cite{Allan563,Umezawa37002} 
However, many studies have also revealed prominent two-gap features, such as shown in Ba$_{0.55}$K$_{0.45}$Fe$_2$As$_2$ (Refs.~\onlinecite{Samuely507, Szabo012503}) and \BaNix up to x = 0.10 (Ref.~\onlinecite{Ren2891}). Furthermore, a wide range of scanning tunneling spectroscopy experiments have observed superconducting gap magnitudes indicative of strong coupling, with $2\Delta/k_BT_c$ ratios far above the BCS weak-coupling expectation of 3.5 (Ref.~\onlinecite{Hoffman124513}). For example, experiments on optimally doped \BaCox have observed a superconducting tunneling gap with coherence peaks corresponding to a much larger (average) single-gap value of $\Delta$ = 6.25~meV, corresponding to $2\Delta/k_BT_c = 5.73$ (Ref.~\onlinecite{Yin097002}).


\section{Angle-Resolved Photoemission Spectroscopy}

ARPES allows for direct visualization of electronic structure of materials and precise characterization of the gapping of occupied states by the superconducting order parameter, and is therefore well-suited for identifying and characterizing the momentum-resolved energy scales of the gap function $\Delta({\bf k})$, in particular for elucidating the gap structure of multi-band superconductors. 

Measurements were performed at $1^3$ end station at BESSY II 
(Helmholtz-Zentrum f\"{u}r Materialen und Energie). Data were recorded from 
freshly cleaved samples. Samples were cleaved in high vacuum at low 
temperatures, exposing mirror-like smooth surfaces.
The Fermi surface map of \BaPt, shown in Fig.~7(a), exhibits large electron pockets at 
the $X$ point and smaller hole pockets at the $\Gamma$ point, which is consistent 
with other electron-doped 122 FeSCs.\cite{Richard124512} Energy momentum cuts presented in Figs.~7(b) and (c) show occupied bands above and below \Tc. Several energy distribution 
curves (EDCs) were measured and are presented in Fig. 7(d)-(g). These curves show the 
consistent appearance of a peak, which is consistent with a gap opening below approximately 17~K on both the electron and hole pockets. For example,  upon cooling below \Tc the appearance of a sharp coherence peak and gapping at the Fermi energy are clearly seen in the EDC curve presented in Fig.~7(d). The width of these features corresponds to approximately a 3~meV gap, which is consistent with the small gap size $\Delta_1$ extracted from fits of the point contact spectroscopy data in Sect.~IV.

Integrated energy distribution curves (IEDCs) were also recorded across several 
cuts at different positions along the electron pocket, as shown in Fig.~7(g). Each 
IEDC has a peak of approximately the same width, confirming that the observed electron pocket gap is isotropic. This is comparable to several other 122 FeSCs studied previously, such as optimally doped \BaKx (Ref.~\onlinecite{Nakayama67002}), \BaCox (Ref.~\onlinecite{Terashima7330}) and LiFeAs (Ref.~\onlinecite{Borisenko67002}), which all show multiple gap sizes with little or no anisotropies.

\begin{figure*}[htbp]
 \includegraphics[width = \textwidth]{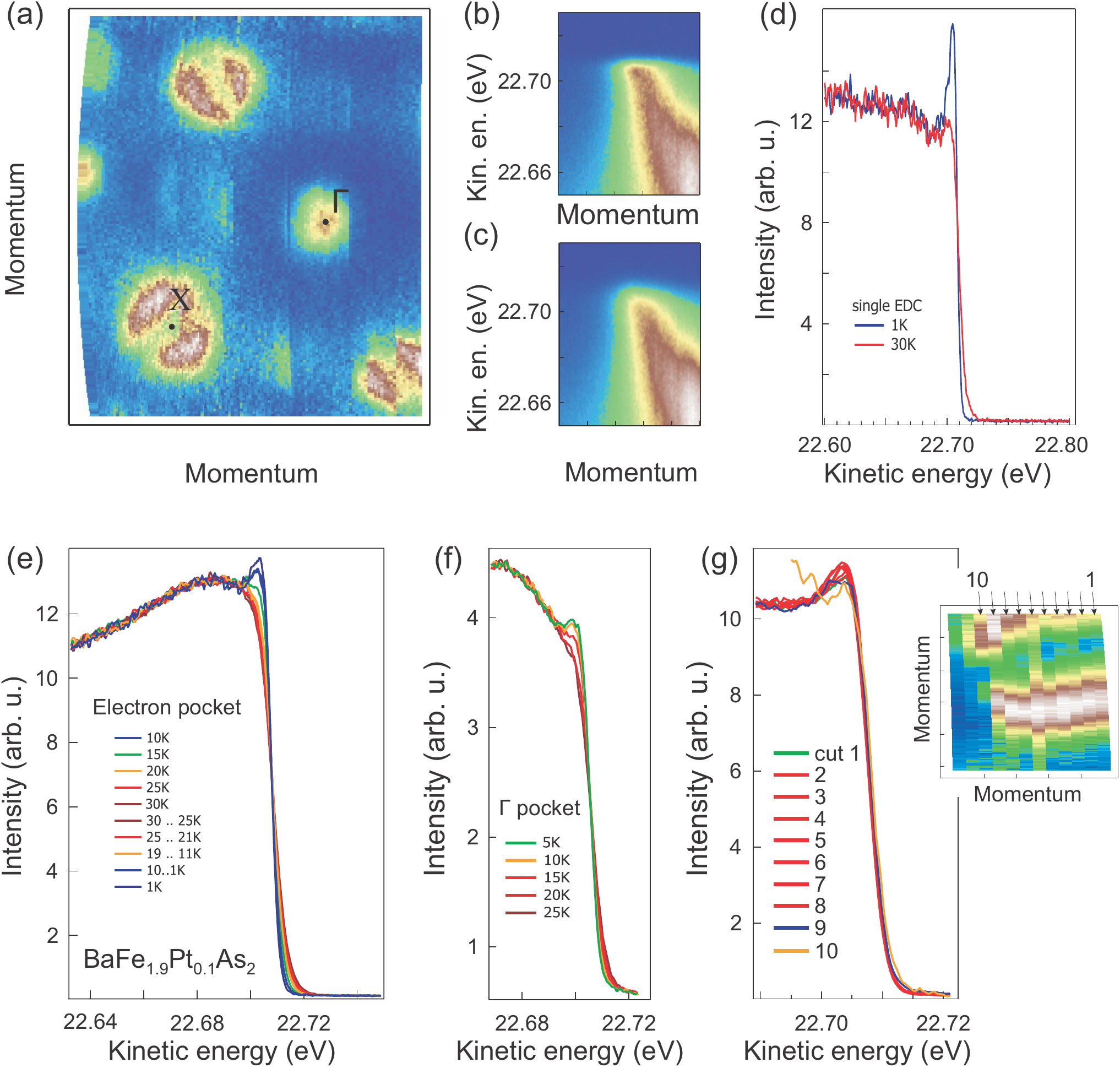}
 \caption{\label{ARPES} ARPES measurements of \BaPtx. (a) Fermi surface map, 
 revealing large electron-like pockets around Brillouin zone corner and small 
 hole-like pockets at center. (b) Energy-momentum cut, passing through the 
 electron pocket recorded below $T_c$. (c) Same cut, recorded above $T_c$. (d) 
 Single energy distribution curve (EDC) recorded above and below \Tc shows 
 appearance of a sharp peak in the superconducting state. (e) Temperature 
 dependence of the integrated EDC (IEDC) recorded from electron pocket for 
 heating up to 30 K and cooling down to 1 K. (f) Temperature dependence of the 
 IEDC for hole pocket. (g) IEDC recorded from different parts of the electron 
 pocket. Positions of cuts are indicated in the mini-map in inset.}
 \end{figure*}


\section{Raman Spectroscopy}

Raman spectra were excited with Kr+ laser lines with 647 and 476 nm wavelengths and 1.2 to 12 mW 
of incident laser power focused into a spot of 50$\times$100 $\mu$m$^2$ on the 
freshly cleaved $ab$-plane crystal surface. The scattered light collected close 
to the backscattered geometry was focused onto 100$\times$240 $\mu$m$^2$ entrance 
slits of a custom triple-stage spectrometer equipped with 1800 lines/mm 
gratings. To record symmetry-resolved Raman spectra of \BaPt we employed both 
linearly- and circularly-polarized light.\cite{Muschler180510} Data were 
collected from 14 total samples at 3 and 30 K. The estimated local heating in the 
laser spot did not exceed 4 K for laser power less than 2~mW.

\begin{figure}[!]
  \centering
 \includegraphics[width = 3.25in]{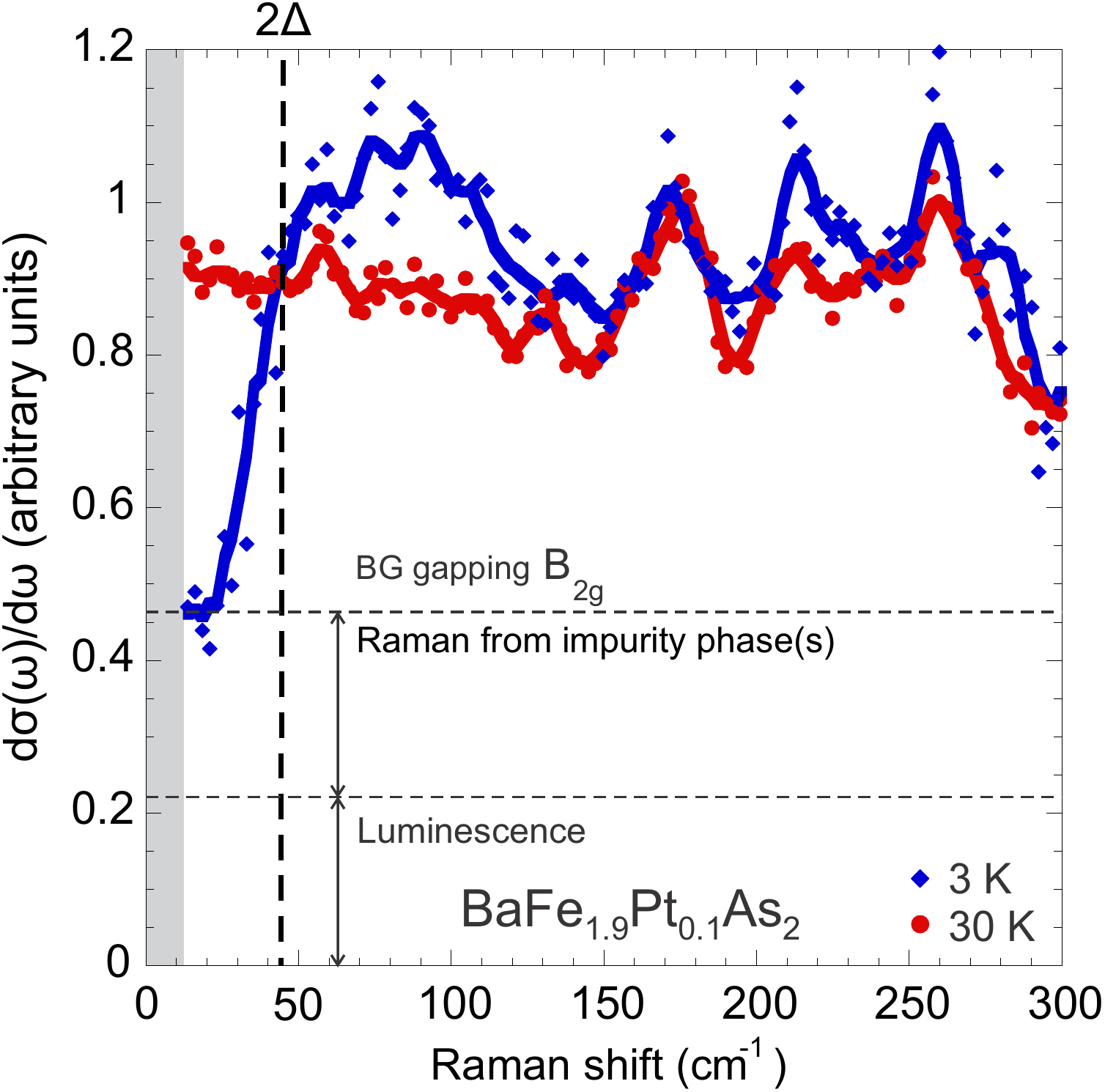}
 \caption{\label{Raman_raw} Raw measured data, containing contributions from 
 both Raman and luminescence channels, for sample S3, obtained with laser 
 excitation of 467 nm. Blue diamonds represent data at 3 K, while red circles 
 represent 30 K data. The superconducting gap is indicated by the vertical 
 dashed line labeled 2$\Delta$. Note that the gray area indicates energies below the 
 cutoff value of the spectrometer.}
\end{figure}

Raw Raman data, $d\sigma(\omega)/d\omega$, normalized 
to the power and laser frequency in the $B_{2g}$ channel are shown in Fig. 8. The plot illustrates the background 
decomposition in the mixed phase doped iron pnictides and highlights the
 contribution of the major superconducting phase to the Raman response. The 
 embedded flux gives rise to laser-induced luminescence. The magnitude of this
 background luminescence (bottom arrow) is evaluated as the minimal polarization contribution 
 for each measured polarization. This background contribution is not related to Raman excitations and 
 must be subracted before $d\sigma(\omega)/d\omega$ is converted into the Raman 
 response. The final spectra are calculated as $\chi^{\prime\prime}$ = ($d\sigma(\omega)/d\omega - 
 BG$)($n(\omega,T)+1$), where $n(\omega)$ is the Bose factor. The resulting 
 response is further decomposed into a sum of contributions from the major 
 superconducting phase and impurity phase(s). These potential impurity phases may not be superconducting or 
may simply have a much lower $T_c$. Based on the flattened section of the experimental data (from the cutoff 
 near 10 cm$^{-1}$ to roughly 25 cm$^{-1}$) at 3 K and general flattened response at 30 K, we 
can assume that contributions from any impurity phase(s) can be well approximated by a 
 constant, much like the luminescence term. Total background (BG) gapping $B_{2g}$ is removed and $d\sigma(\omega)/d\omega$ 
 is converted to the Raman response, $\chi_{major}^{\prime\prime}$ = ($d\sigma(\omega)/d\omega - 
 BG$)($n(\omega,T)+1$).

The normal state response (red line) is essentially flat down to the cutoff value of the 
spectrometer. The superconducting response (blue line) exhibits a broad peak 
around 80-90 cm$^{-1}$. This represents the strong feature clearly seen in previous Raman 
studies \cite{Muschler180510, Chauviere180521, Sugai140504} and a threshold near 
45 cm$^{-1}$ marked by a black dashed line. This response flattens out below approximately 25 
cm$^{-1}$. The flattened background (uppermost black dashed line) is a result of  
sample impurity content, as observed elsewhere. All 14 samples measured by Raman spectroscopy showed 
non-vanishing backgrounds in which amplitude depends on spot position within the 
cleave and laser excitation wavelength. 

We focus on properties of a major superconducting phase removing the contribution from the 
impurity phase(s) as it is described above. The measured electronic Raman response, $\chi^{\prime\prime}(\omega)$, in the $B_{2g}$ 
and $A_{1g}$ channels is depicted in Figs. 9(a) and (b), respectively. The data in 
the SC state exhibit a threshold around 45-50 cm$^{-1}$ (vertical black dashed 
line, labelled as 2$\Delta$), clearly seen in both $B_{2g}$ and $A_{1g}$ 
polarizations. This is a fundamental gap, also seen in ARPES and point-contact spectroscopy. The value of $\Delta = 3 
\pm 0.3$~meV was confirmed for multiple \BaPt samples. 

\begin{figure}[!]
  \centering
 \includegraphics[width = 3.25in]{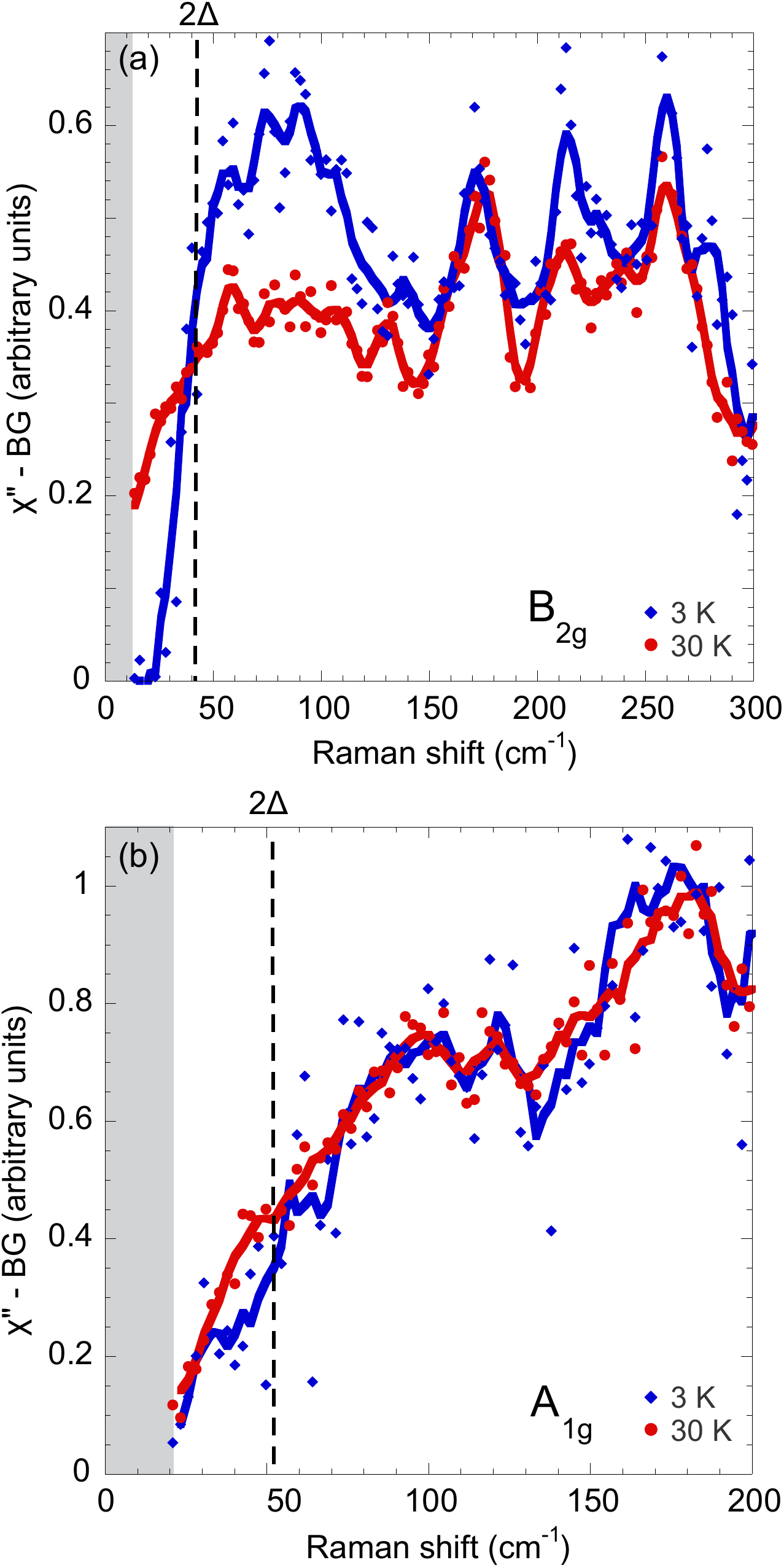}
 \caption{\label{Raman_S3_B2g_A1g} Electronic Raman response (background removed) 
 of sample S3, obtained with 476 nm laser excitation in (a) $B_{2g}$ and (b) $A_{1g}$ 
 channels. Blue (red) dots show superconducting (normal state) response averaged 
 over four (two) experimental scans. Both thick lines represent 5-point smoothed 
 data. Black dashed lines mark the energies corresponding to the 
 superconducting gap observed, 2$\Delta$. The gray area again indicates areas 
 below the cutoff energy of the spectrometer.}
\end{figure}


\section{Discussion}

In order to understand these results in context it is instructive to recall the 
conclusions drawn from similar experiments on Co-substituted BaFe$_2$As$_2$. Results from multiple studies on optimally doped \BaCox tend to confirm 
a similar nodeless gap structure. Comprehensive analysis of the phase diagram by 
Reid \etal\cite{Reid064501}~looked at the evolution of gap structure with increasing Co concentration via thermal conductivity. 
Based on the presence or absence of residual thermal conductivity \kzeroT and its
evolution in field, it was shown that line nodes are present in 
under- and overdoped \BaCox but that optimally doped $x$ = 0.148 samples were 
nodeless.

Daghero \etal~measured several 
PCS spectra for overdoped samples with $x$ = 0.2 (Ref.~\onlinecite{Daghero124509}). Current was injected in the $ab$-plane and parallel to the $c$-direction using the 
soft point contact method, in which a small amount of conductive paint is used to create 
the N-S junction rather than pressing a sharpened tip 
into the SC sample. Attempts to fit to BTK theory showed a better fit for a 
two-gap isotropic $s$-wave model than for a single nodeless gap. These fit curves included estimated gap 
values of $\Delta$$_1$ = 3.8-5.2 meV and $\Delta$$_2$ = 8.2-10.9 meV. The 
authors also reported features in the measured $dI/dV$ curves at higher bias voltage that deviated from the BTK fit 
which were attributed to strong electron-boson coupling. PCS measurements were 
also performed by Samuely \etal~with sharpened Pt tips pressed into freshly 
cleaved optimally doped ($x$ = 0.14) crystals.\cite{Samuely507} BTK fits to the measured dI/dV 
curves showed only a single isotropic gap of magnitude $\Delta$ = 5-6 meV 
suggesting that if multiple gaps are present they are close together in 
magnitude. Given the significant differences in gap structure observed through 
thermal conductivity across the \BaCox phase diagram, it is possible that the stark difference in these results 
may be attributed to their different Co concentrations.

Terashima \etal~examined the gap structure of $x$ = 0.15 samples using ARPES and 
found strong evidence for two isotropic gaps.\cite{Terashima7330} Energy 
distribution curves showed two gaps with $\Delta$$_1$ = 5 $\pm$ 1 meV on the 
electron pocket and $\Delta$$_2$ = 7 $\pm$ 1 meV on the hole pocket. EDCs measured over a range of 
angles shows that both gaps are isotropic within error. 

Muschler \etal~performed Raman spectroscopy measurements on \BaCox samples with optimal doping 
($x$ = 0.122) and slight overdoping ($x$ = 0.17).\cite{Muschler180510} Based on the 
strong low-temperature shift in the $B_{2g}$ spectrum which varies as $\sqrt{\Omega}$ 
the authors propose an $s$-wave state with accidental nodes in the optimally doped 
sample. Slight overdoping resulted in a $B_{2g}$ peak with greatly diminished 
amplitude suggesting that the gap is strongly effected by doping and disorder 
(and potentially sample quality).

The most agreeable conclusion regarding the gap structure 
of \BaCox seems to be that of an $s$-wave gap which is nodeless at optimal doping 
with nodes appearing at higher and lower Co concentrations which do not appear 
to be imposed by symmetry and which are present away from the $ab$-plane. However, 
some results appear to contradict each other. For example, PCS studies 
consistently argue in favor of an isotropic gap while Raman spectra at multiple 
dopings suggest the presence of nodes. Some have suggested reconstruction of the 
gap at the surface as a means to explain differences between measurements, but 
this alone does not reconcile disagreements between thermal conductivity, PCS, 
and Raman spectroscopy, all of which are bulk probes. Another possible 
explanation is differences in crystal quality, and thus impurity scattering, 
which has been proposed by Muschler and others.

Compared to the Co-substituted system, our results on \BaPt offer a more comprehensive conclusion of an isotropic gap structure as 
all of the samples used were made in the same batch and therefore are expected to have 
similar impurity concentrations and defect structures. Furthermore, our results all support the conclusion of 
at least one gap with a consistent magnitude of approximately 3~meV, and all experiments point to an isotropic $s$-wave gap structure.

Looking at the Co- and Pt-doping comparison more directly, both optimally doped materials lack a residual thermal conductivity 
at zero field, indicating a fully-gapped superconducting order parameter. For the Co compound, a nonzero \kzeroT 
emerges at small magnetic fields, while for the Pt compound it remains zero at all fields 
observed. While the thermal conductivity of each material could only be observed up to a small fraction of 
$H_{c2}$, this may suggest subtle differences in gap morphology. PCS measurements on 
both yielded BTK fits to nodeless $s$-wave models. Results from our study 
with optimal Pt doping seem to compare more directly with those of overdoped 
\BaCox rather than optimally doped. Both point to a two-gap structure with 
features present in the spectrum that do not perfectly match the BTK fit, which may indicate strong electron-boson coupling in both materials. ARPES measurements on optimally doped 
samples in both systems showed no variation in gap magnitude as a function of 
angle.\cite{Terashima7330} In the Co-doped compound, two gaps were consistently seen, while only one was observed in our study of BaFe$_1$$_.$$_9$Pt$_0$$_.$$_1$As$_2$. Finally, Raman spectra on both compounds feature 
low temperature enhancements in the $A_{1g}$ and $B_{2g}$ channels at 
positions which agreed with other reported values of gap size.

While we found no evidence of nodes for optimally doped BaFe$_1$$_.$$_9$Pt$_0$$_.$$_1$As$_2$, 
under- and overdoped compounds in the \BaPtx system have yet to be explored as 
far as gap structure is concerned. It has yet to be seen whether similar nodes 
would appear or if the Pt-substituted system would exhibit significant 
differences from Co substitution. Therefore, future investigation of the entire 
\BaPtx phase diagram is certainly warranted.

Our results on \BaPt can also be used to draw interesting parallels with the multiband superconductor LiFeAs. Both 
compounds present no evidence of quasiparticle excitations in thermal 
conductivity measurements at low magnetic fields, and show minimal gap anisotropy in gap structure as determined by ARPES measurements.
Both also exhibit comparable small and large gap magnitudes as extracted from PCS measurements and ARPES measurements in the case of LiFeAs, with a 2$\Delta$/$k_B$\Tc ratio of approximately 2 for the smaller gap in each system. The multi-gap nature of both materials would seem to be in disagreement with the thermal conductivity observations, but 
can be reconciled if one assumes the coherence length is similar between the two 
gaps. In our PCS observations of BaFe$_1$$_.$$_9$Pt$_0$$_.$$_1$As$_2$, the large discontinuity between Z values for the two gaps is consistent with large differences between Fermi velocities, which means that $v_F/\Delta$ could be comparable for the two gaps as suggested for LiFeAs.\cite{Tanatar054507}

\section{Conclusion}

In conclusion, we have used four measurement techniques to probe the superconducting gap structure in single-crystal samples of BaFe$_1$$_.$$_9$Pt$_0$$_.$$_1$As$_2$, finding reliable evidence for an isotropic one- or possibly two-gap $s$-wave model. 
Thermal conductivity measurements show no sign of low energy quasiparticle excitations even at relatively high magnetic fields, with behavior comparable to other isotropic, single-gapped $s$-wave superconductors. While there is no indication of a reduced energy gap, the presence of a second gap cannot be excluded based on the small relative field range that was studied. 

Conductivity spectra measured by point-contact Andreev reflection spectroscopy exhibit sharp enhancements and notable suppression of $dI/dV$ at lower and higher bias, respectively, which suggests the 
presence of two gaps. Fitting to an isotropic two-gap Blonder-Tinkham-Klapwijk model results in gap size estimates of 2.5~meV and 7.0~meV, corresponding to features in the spectra that have been replicated in several crystals from the same batch. 

Angle-resolved photoemission spectroscopy measurements observe an isotropic gap on both electron and hole bands with a magnitude of approximately 3~meV. Finally, Raman spectroscopy revealed excitations in the superconducting state in both $A_{1g}$ and $B_{2g}$ channels whose energy scales correspond with the 3~meV gap magnitude observed in other measurements.

Overall, we conclude that the optimally doped iron-based superconductor \BaPt has an isotropic, fully gapped superconducting order parameter with no nodes or deep minima, possibly with band-dependent energy scales of order 3~meV and 7~meV. Combining several experimental probes together rules out several extrinsic parameters, allowing for further elucidation of the peculiar multi-band nature and pairing mechanism of this and other iron-based superconductors.

$ $

The authors would like to acknowledge R. L. Greene and I. Takeuchi for 
valuable discussion and X. Zhang for experimental and analytical 
assistance. Research at the University of Maryland was supported by AFOSR-MURI 
Grant No. FA9550-09-1-0603 and NSF-CAREER Grant No. DMR-0952716. Work at 
Sherbrooke was supported by the Canadian Institute for Advanced Research and 
funded by NSERC, CFI, FRQNT, and a Canada Research Chair. Work at Rutgers 
was supported by the US Department of Energy, Office of Basic Energy Sciences, 
Division of Materials Sciences and Engineering under Award DE-SC005463.



\begin{thebibliography}{999}
\bibitem{PCSfootnote} Features in the conductance spectra associated with the onset of superconductivity in sample S2 appear to emerge only below 18~K, which is lower than the nominal $T_c$ value of 23~K measured by resistivity, susceptibility and heat capacity  (Ref.~\onlinecite{Saha072204}). This may be due to critical current effects or possibly a departure from the diffusive Sharvin regime at higher temperatures.\cite{Sheet134507}
\end{thebibliography}
\end{document}